\documentclass[aps,pre,nofootinbib,showpacs,twocolumn]{revtex4-1}%
\usepackage{amssymb,latexsym,mathrsfs}
\usepackage{amsfonts}
\usepackage{mathrsfs}
\usepackage{amsmath}
\usepackage{graphicx}

\newcommand{\beq}{\begin{equation}}
\newcommand{\eeq}{\end{equation}}
\newcommand{\beqn}{\begin{eqnarray}}
\newcommand{\eeqn}{\end{eqnarray}}
\newcommand{\bearr}{\begin{array}}
\newcommand{\enarr}{\end{array}}

\def\bea{\begin{eqnarray}}
\def\eea{\end{eqnarray}}
\def\ba{\begin{array}}
\def\ea{\end{array}}
\def\n{\nonumber}

\def\pr{\shortparallel}

\begin{document}
\title{Phase Transition in an Exactly Solvable Extinction Model} 

\author{ Debarshee  Bagchi}
\email[E-mail address: ]{debarshee.bagchi@saha.ac.in}
\author{P. K. Mohanty}
\email[E-mail address: ]{pk.mohanty@saha.ac.in}
\affiliation{Theoretical Condensed Matter Physics Division, Saha Institute of Nuclear Physics,
1/AF Bidhan Nagar, Kolkata 700064, India.}
\date{\today}

\vskip 2.cm

\begin{abstract}
We  introduce a model of biological evolution where species evolve in response to 
biotic interactions and a  fluctuating environmental stress. 
The species may either become extinct or mutate to acquire a new fitness value when 
the {\it effective} stress level is greater than their individual fitness. The model exhibits 
a  phase transition to a completely extinct phase as the environmental stress  
or the mutation rate is varied. We discuss the generic conditions for which this 
transition is continuous. The model is exactly solvable and the  critical behavior 
is characterized by an unusual dynamic exponent $z=1/3.$ 
Apart from predicting large scale evolution, the model can be applied 
to understand the trends in the available fossil data.
\end{abstract}
\maketitle

An important question that intrigues both scientists and non-scientists alike,
is that of the origin and evolution of life on this planet. Where did all the
diverse variety of species on earth come from? 
Why did some of the species which initially existed, got wiped out
while others survived? What are the factors influencing the process of mass extinction?
These have emerged as a multi-disciplinary field of research over the last few decades,
attracting attention of researchers from various branches of science.
The phenomenon of species extinction is an equally important event, and is an
inextricable part of the evolution process. In recent times, several models of 
evolution and extinction have been proposed.
Some early evolution models have considered evolutionary dynamics of 
interacting species on a rugged \textit{fitness landscape} \cite{Wright,NK}.
In these models, under repeated mutation and selection \textit{fit} species tend to
{\it climb up} on the fitness landscape until a local maximum or peak is reached. 
The landscape may also be coevolving with the evolution of the species.
A pioneering work by Bak and Sneppen \cite{Bak} modeled the extinction of
coevolving species on a fitness landscape as a self-organized critical 
(SOC) process. Here, coevolution of species can trigger \textit{coevolutionary
avalanches} of extinction events and the avalanche
size is distributed algebraically with an universal exponent 
$\tau$ in the range $1 \le \tau \le \frac 32$\cite{review}.
However, a major drawback of these models is that they completely 
ignore the role of the environmental stresses e.g., climatic, 
geological or exogenous stresses\cite{review}.
Recently, Newman\cite{Newman} has proposed a model considering only 
environmental stresses as the cause of large scale extinction, 
which can explain the trends in the available fossil records\cite{data}.

In this article, we study a model of biological evolution taking into account
both external stresses as well as biotic interactions between species
as contributory factors for species extinction.
Here, less fit species either become extinct or they mutate, 
due to changes in environmental stress level. Biotic interaction is incorporated as
\textit{cooperativity} among the species, which has not been
considered in any of the models mentioned above. 
It is well known that cooperativity is an important factor for 
proper functioning of every ecosystem which arises from the
interdependence between species, e.g., \textit{via} the food web\cite{food-web}. 
Such interactions at the level of individual species are in fact important 
in modeling evolution and extinction on ecological time scales, 
as has been argued in Ref.~\cite{DC-etal}.
%

We show that, a phase transition leading to a complete extinction of the
species may occur  in this model, as the environmental stress is increased.
Generally such a transition occurs discontinuously and only under specific conditions 
it becomes continuous. For the continuous transition, the critical value of the stress, 
the the critical exponents  $(\alpha, \beta, \nu_\pr, \nu_\perp, z)$ 
and the scaling functions can be calculated analytically. The critical behavior 
of the system is found to be robust against the variation of mutation rate and the
fluctuations in stress.

Let us now describe the model in detail. We consider an ecosystem consisting 
of $N$ different species and let $x_i$ ($i = 1,2,...,N$) denote the fitness of
the $i^{th}$ species. The fitness $x_i$ is drawn from a fitness distribution
$\mathcal{F}(x)$.
The species are subjected to a fluctuating environmental stress $\mathcal{S}$ which is drawn 
from a distribution $\mathcal{D}(\mathcal{S})$. However, due to the presence of cooperativity, 
the existing species in the ecosystem experience an \textit{effective} stress 
$ s = {\mathcal{S}}/{\cal C}$, where ${\cal C}$ is the measure of cooperativity among the species.
Thus, for a given environmental stress $\mathcal{S}$, the effective stress
is smaller when cooperativity among species is large and vice-versa.
In general, ${\cal C}(N_t)$ depends on the existing number of species $N_t$ at time $t$.
If the number of existing species is very large, they compete for resources e.g., food, habitat etc.
and their cooperativity decreases. Thus ${\cal C}(N_t)$ is expected to be a decreasing
function for large $N_t$. However, for very small $N_t$, the species 
tend to cooperate for survival. In this regime, the resulting  cooperativity
${\cal C}(N_t)$ is an increasing function.

Extinction of species in our model occurs as follows.
At any instant $t$, species whose fitness $x_i$ is numerically
smaller than the effective stress $ s_t = S_t/ {\cal C}(N_t)$, either
become extinct with rate $p,$ or mutate with rate $(1-p)$ to a new fitness
$x_i$ randomly drawn from the distribution $\mathcal{F}(x)$. 
Note that the renewed fitness value may be either same, greater, 
or smaller as compared to the current value of the species fitness, 
and correspond to a neutral, favorable or an harmful mutation respectively.

Since in our model the extinct species are not repopulated by new ones, 
the number of existing species $N_t$, and therefore, the density $\rho_t = N_t/N$, 
can only decrease with time. The new effective stress is then 
$s_{t+1}=  S_{t+1}/ {\cal C}(N_{t+1}),$ where, $S_{t+1}$ is a new
environmental stress value drawn from ${\cal D (S)}.$ The dynamics
stops when either all the the surviving species are fit, or none
of them survive under the applied stress. Therefore, depending on
the environmental stress there is a possibility of complete extinction
of the species. This transition, from a phase with finite population to 
complete extinction, may occur continuously or abruptly as 
$ \langle S \rangle = \int S{\cal D(S})dS $ is increased.
We will characterize this transition in detail considering 
${\sigma} = \langle S \rangle /N$ and $p$ as control parameters,   
and  provide a phase diagram in the $p$ - $\sigma$ plane.  

First, let us  assume that, (i) the resources are infinite and 
therefore, the cooperativity ${\cal C}(N_t)$ increases monotonically,
say linearly, with $N_t$, $i.e.$, ${\cal C}(N_t) = N_t$ and 
(ii) the environmental stress $\mathcal{S}$ do not fluctuate in time, $i.e.$,  
$\mathcal{D} (\mathcal{S}) = \delta (\mathcal{S}-\mathcal{S}_a)$.
The more general cases, like different functional forms of ${\cal C}(N_t)$ and 
effects due to fluctuation in the stress will be discussed at the end.

\begin{figure}[ht]
\centerline
{
\includegraphics[width=8.0cm,height=6.0cm,angle=0]{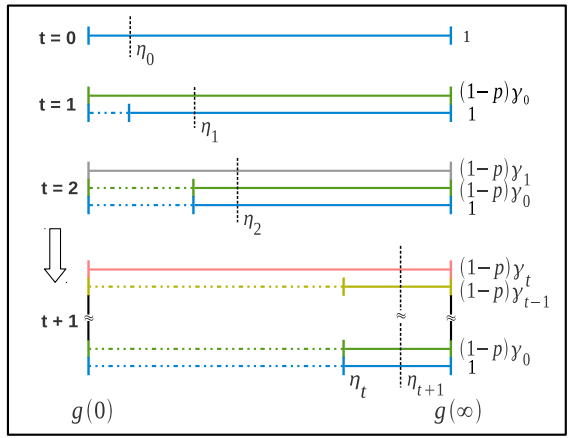}
}
\caption{(Color online)
Schematic description of the removal and renewal events. The $x$-axis represents the cumulative distribution, which 
extends from $g(0)=0$ to $g(\infty)=1.$  At a given time $t$, the species which are surviving (extinct) are denoted by  solid 
(dotted) horizontal lines and a marker $\eta_t= g(\sigma/\rho_t)$  is used to indicate the point below which
the species are  considered {\it unfit}.  Of these surviving  {\it unfit}  species  (their density is $\gamma_t$), 
$1-p$ fraction will mutate to have new fitness values extending over the whole range, increasing  the  
density of surviving species  at $t+1$ by $(1-p)\gamma_t$. Effectively, at $t + 1$  all the solid lines 
up to $\eta_t$ are removed and one additional solid line with weight $(1-p)\gamma_t$ is drawn.
}
\label{fig:cartoon}
\end{figure}

Starting from  a initial density $\rho_0 = 1$, the density of existing species evolves  as,

\vskip-0.75cm
\beq
\rho_{t+1} = \rho_t - p \gamma_t = \rho_0 - p \sum_{t'=0}^t \gamma_{t'}
\label{eq:evol-eqn-p}
\eeq
where, $\gamma_t$ is the density of the {\it unfit} species at time $t$, 
$i.e.$  $N \gamma_t$ is the number of existing species with fitness value
smaller than $ {\cal S}_a/N_t = \sigma/\rho_t $. Clearly, $p$ fraction
of $N \gamma_t$ species becomes extinct, and $1-p$ fraction undergo
mutation, acquiring new fitness value spanning the entire range of
distribution function ${\cal F}(x)$.

Effectively, the above dynamics amounts to the removal of all the $N \gamma_t$ 
unfit species from the system at time $t$, and introducing $N (1-p)\gamma_t$
species with renewed fitness value.
The time sequence of such \textit{removal} and \textit{renewal} 
events are schematically shown
in Fig.~\ref{fig:cartoon}. The solid horizontal lines extending
from $g(0) = 0$ to $g(\infty) = 1$ represent the density of species which
underwent mutation at the previous time instant and is equal to
$(1-p)\gamma_{t}$; $g(x)= \int_0^x {\cal F}(x') dx'$ is the 
cumulative distribution function.
The dashed lines correspond to the species which have been
removed from the system.
Utilizing this, it is easy to calculate the density of unfit species $\gamma_{t+1}$
and is equal to the weighted sum of the length of solid lines upto $\eta_t$, with the
weight factors $(1-p)\gamma_{t}$, where,

\vskip-0.5cm
$$
\eta_t = \int_0^{{\sigma}/{\rho_t}} {\cal F}(x) dx = g (\sigma/\rho_t)
$$
as shown in Fig.~(\ref{fig:cartoon}). Thus, 

\vskip-0.5cm
\beqn
\gamma_{t+1} = (\eta_{t+1} - \eta_t)\left[1 + (1-p)\sum_{t' = 0}^{t-1}\gamma_{t'} \right] + (1-p)\gamma_{t}\eta_{t+1} \n \\
\label{eq:gamma1}
\eeqn
Eqs.~(\ref{eq:evol-eqn-p}) and  (\ref{eq:gamma1}) describe the dynamical rules of our model and can be recast 
into a simple form using $\displaystyle{ \pi_{t+1} = \sum_{t' = 0}^{t} \gamma_{t'}}$;

\vskip-0.5cm
\beqn
\pi_{t+1} &=& \eta_t[1 + (1-p)\pi_t ] \n \\
\rho_{t+1} = \rho_0 - p \pi_{t+1} &=& 1- p \eta_t[1 + (1-p)\pi_t ]
\label{eq:itr-eqns}
\eeqn

When mutation is absent $(p=1)$, this set of equations 
Eq.~(\ref{eq:itr-eqns}) reduces to $\rho_{t+1} = 1 - \eta_t$.
This special case of our model, with ${\cal C} (N_t)= N_t$ and  
${\cal D (S)}  = \delta( {\cal S- S}_a)$ has been studied earlier 
\cite{FBM} as the \textit{democratic Fiber Bundle Model} (dFBM)
in the context of failure processes.
In the dFBM model, a heavy load weighing ${\cal S}_a$ hangs
from a rigid anchor by a bundle of $N$ elastic fibers, each having
a certain breaking strength $x_i$. Initially, all the
fibers are intact and share the applied load equally, 
each  experiencing an effective load ${\cal S}_a/N.$  At each time step,
weaker fibers (fibers with strength less than the effective load) fail,
and the load is re-shared equally among the remaining intact fibers.
Thus the effective load per fiber increases creating an avalanche of
failure events. If the initial load $S_a$ is low, this process reaches 
a stationary state with some intact fiber which eventually 
support the load, whereas a complete failure occurs for high $S_a.$ 
Thus, at some critical value of $S_a$ the dFBM model exhibits 
a breakdown transition  which  may be  abrupt or continuous\cite{PRL3}. 

Coming back to the general case $p \neq 1$, let us first compute the fixed points of 
Eq.~(\ref{eq:itr-eqns}), where   $\pi_{t+1} = \pi_t = \pi^*$ and  
$\rho_{t+1} = \rho_t =\rho^*$. Thus, we have $\pi^*=  g(\sigma/\rho^*)[1 + (1-p)\pi^*]$
and
\vskip-0.75cm
\beqn
\rho^* = 1- p \pi^*=\frac{1-g(\sigma/\rho^*)}{1-(1-p)g(\sigma/\rho^*)}.
\label{eq:fxd-pt}
\eeqn
This equation may have multiple  solutions
for $\rho^*$; the largest among them $\rho_s={\rm Max}(\rho^*)$ is stable since the right 
hand  side of Eq.~(\ref{eq:fxd-pt}) is  a non-decreasing function of $\rho^*$ bounded
in the range $(0,1).$ 
Therefore, starting from the  initial density $\rho_0=1$, the density decreases and 
eventually approaches a  stationary value $\rho_s.$

\begin{figure}[h]
\centerline
{
\includegraphics[width=3.5cm,height=4.2cm,angle=270]{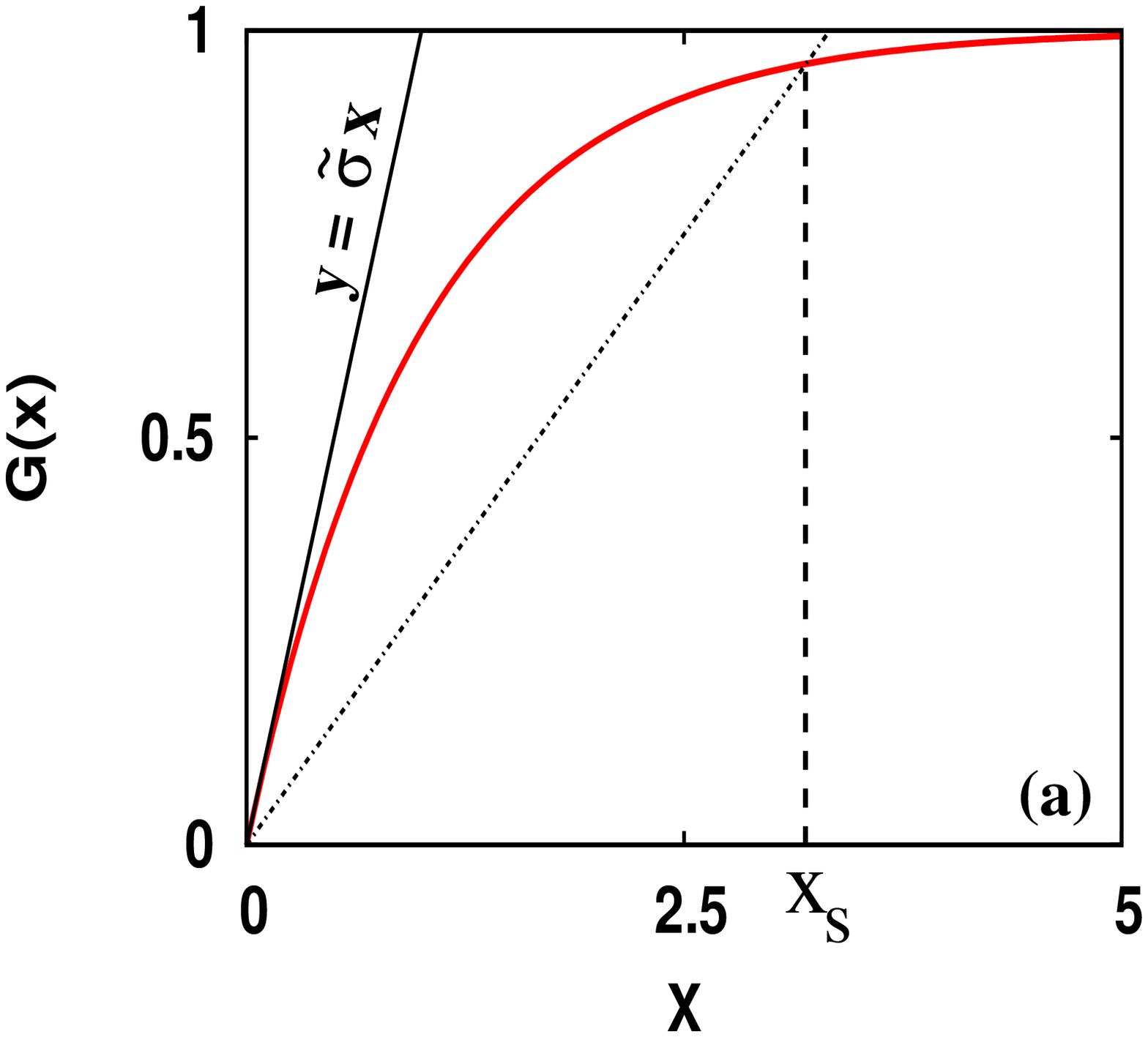}\hskip-0.2cm
\includegraphics[width=3.5cm,height=4.2cm,angle=270]{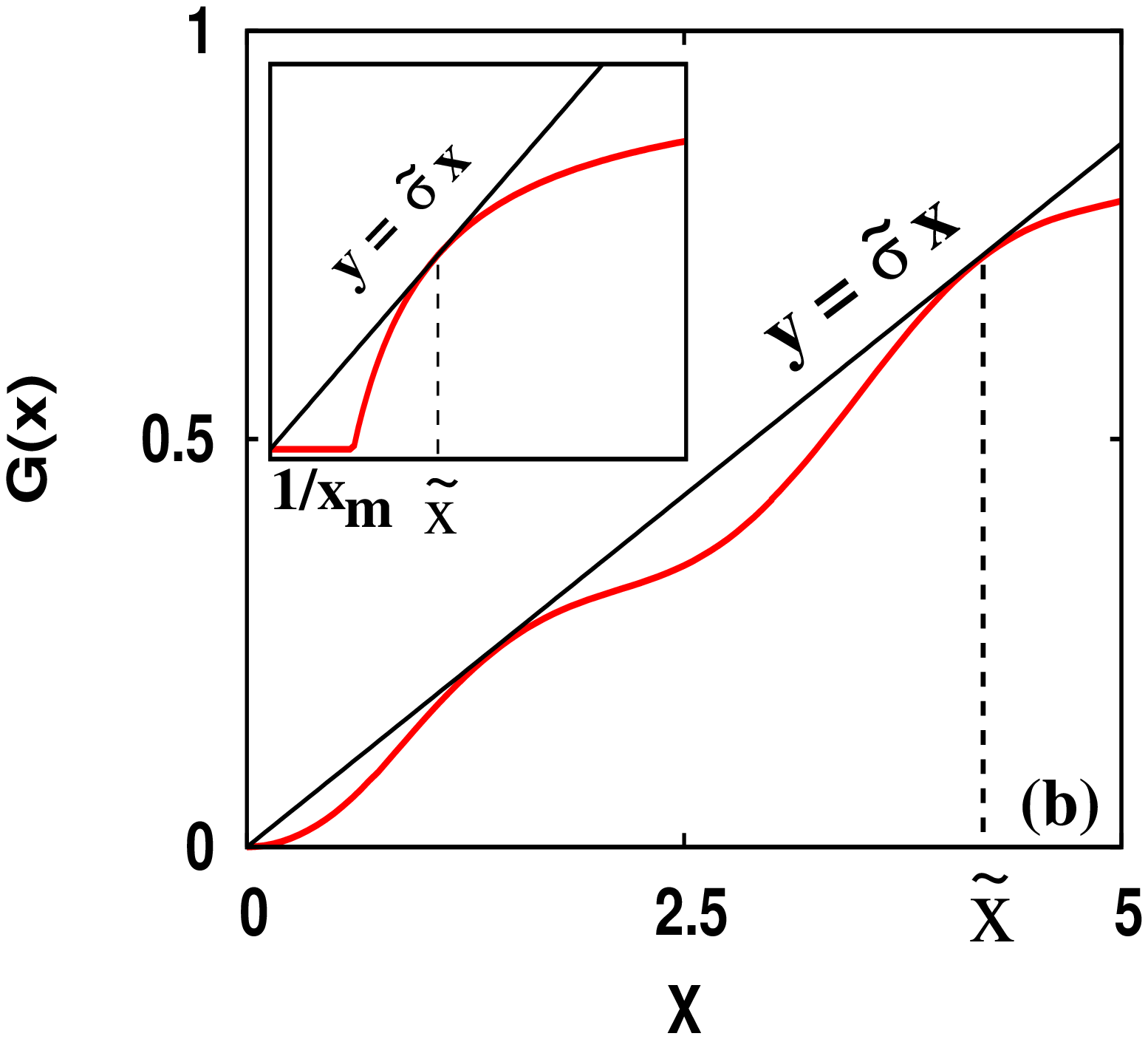}
}
\caption{(Color online) (a) A typical $G({\tt x})$, bounded from above by $y=\tilde \sigma x$  where $\tilde\sigma = G'(0)$,
results in a continuous extinction transition with $\rho_s= \sigma {\tt x}_s.$ (b) For generic $G({\tt x})$ 
the transition is discontinuous;  (inset) $G({\tt x})$ for a bounded  fitness distribution (see text).}
\label{fig:Gx}
\end{figure}

The steady state density $\rho_s$ is the rightmost intersection point of the curves
\vskip-0.55cm
\beq
y = \sigma {\tt x}~~~{\rm and}~~~y = G({\tt x})=\frac{1-g(1/{\tt x})}{1-(1-p)g(1/{\tt x})},
\label{eq:curves}
\eeq
%
%
where ${\tt x}$ replaces $\rho^*/ \sigma$ in Eq.~(\ref{eq:fxd-pt}) and
$G({\tt x})$ is a non-decreasing function with 
$G(0)= 0$ and $G(\infty)= 1$. Clearly, these two curves
intersect at ${\tt x}=0$ for all values of $\sigma$.
If the curves intersect at other points (${\tt x} > 0$), the rightmost
one, (say) ${\tt x}_s$  correspond to the steady state density
$\rho_s = \sigma {\tt x}_s.$
This is described schematically in Fig.~\ref{fig:Gx}a. Existence of a solution 
$\rho_s>0$ indicates that the system is in a non-extinct phase while complete extinction 
occurs when the only solution is $\rho_s=0$.

The transition point between the two phases and the order of the transition 
can be  determined from the  behavior of $G({\tt x}).$  Consider that $G({\tt x})$ is
bounded from above by the a line $y = \tilde \sigma {\tt x}$ such that 
$G({\tt x}) \leq  \tilde \sigma {\tt x}~~ \forall~~ {\tt x}.$  Let this line
be a tangent of $G({\tt x})$ at some ${\tt x}=\tilde {\tt x}$.
If the line is a tangent at multiple points (see Fig.~\ref{fig:Gx}b), the 
rightmost one will be $\tilde {\tt x}.$
Clearly  the system is in  (non-)extinct phase for any $\sigma$ (smaller) greater than 
$\tilde \sigma.$ So, at the transition point $\sigma_c= \tilde \sigma$ the order parameter is
$\rho_s = \tilde \sigma \tilde {\tt x}.$
The transition will be discontinuous when  $\tilde {\tt x}$ is nonzero,
since in this case  the order parameter has a nonzero value at the critical point
(see Fig.~\ref{fig:Gx}b). A continuous transition occurs if $\tilde {\tt x}=0$,
and the critical point is, therefore, $\sigma_c=\tilde \sigma = G'(0).$  This implies that the transition
is continuous only if the tangent line of $G({\tt x})$ at ${\tt x} = 0$ bounds the
curve from above, i.e., when $G({\tt x})/ {\tt x} \leq G'(0)~\forall~{\tt x}$ (see Fig.~\ref{fig:Gx}a).

It may be mentioned here that the extinction transition will be discontinuous if  
$\mathcal{F}(x)$ has a  cut-off (say) at $x_m$, such that $\mathcal{F}(x > x_m) = 0.$
This implies that $g(x > x_m) =  1$ and therefore, $G({\tt x} < 1/ x_m) = 0$. 
A typical form of such a function,  with $x_m=1$,  is shown in Fig.~\ref{fig:Gx}b(inset). 
In this case, the transition will  occur discontinuously as $\tilde {\tt x} \neq 0.$  

In the following, we discuss the continuous phase transition in detail.
Since the critical point $\sigma_c= G'(0),$  is related to $g({\tt x})$
and its derivative as ${\tt x}\to \infty$, we expand $g(x)$ as,  \vskip-0.6cm
\beqn
g(x) &=& 1 - a(1/x)^{\lambda}  + \dots 
\label{eq:gx}
\eeqn
%
where $a$ is a positive constant. Thus for $\lambda < 1$, the critical point $\sigma_c = G'(0) = \infty$
and hence complete  extinction can never happen.  
Again, for  $\lambda > 1$ the transition is  discontinuous as in this case $G'(0)=0.$ 
Thus, the only non trivial extinction phase transition occurs  for $\lambda = 1$  where the critical 
point $\sigma_c = a/p$ is  nonzero and finite.  Note that the higher order terms in Eq.~(\ref{eq:gx}) 
do not play any role in determining the critical point.


Let us now characterize the phase transition in terms of the critical exponents and the scaling
functions.
From Eq. (4), it is evident that for small $\varepsilon= \sigma_c - \sigma$, 
the  order parameter $\rho_s=  (\varepsilon/\sigma_c) ^\beta$, with  $\beta=1.$
The scaling functions can be derived from the dynamics of the  model near the critical
point, which can be approximated, following Eq.~(\ref{eq:fxd-pt}), by 

\vskip-0.6cm
\beqn
\sigma \rho_{t+1} &=& \frac{1-g(1/\rho_t)}{1-(1-p)g(1/\rho_t)}.
\label{eq:fxd-pt-itr}
\eeqn
\vskip-0.25cm

Taking the continuum  limit and retaining terms upto second order in $\rho$, we have 
\vskip-0.6cm
\beqn
(a/p - \varepsilon)\frac{d\rho}{dt} & = & \varepsilon \rho - \frac{a^2(1-p)}{p^2} \rho^2,
\label{eq:eps-t}
\eeqn
\vskip-0.25cm
where $\sigma_c = a/p$. Close to the critical point $(\varepsilon \rightarrow 0)$,
by rescaling the variable as $\varrho = \varepsilon^{-1} \rho$ and ${\tt t} = \varepsilon t$,
Eq.~(\ref{eq:eps-t}) can be written in a scaling form as
$\frac ap\frac{d\varrho}{d{\tt t}} = \varrho - \frac{a^2(1-p)}{p^2} \varrho^2$.
The formal solution of this rescaled differential equation can be written
in one of the following forms;
\vskip-0.6cm
\beqn
\rho(\varepsilon, t)
=\begin{cases}
& \displaystyle t^{-\alpha} f_{\alpha}(\varepsilon^{\nu_\shortparallel} t) \\
& \displaystyle \varepsilon^{\beta} f_{\beta}(\varepsilon^{\nu_\shortparallel} t), 
\end{cases}
\label{eq:scaling-eqn}
\eeqn
where, the exponents $\alpha = 1 = \nu_\shortparallel.$ From the scaling 
relation $\beta=\alpha \nu_\shortparallel$, we again get $\beta=1$.
The functions $f_{\alpha}(x)$ and $f_{\beta}(x)$ can be obtained by solving Eq.~(\ref{eq:eps-t})
with boundary condition $\rho(\varepsilon, 0) = 1$ which gives,

\vskip-0.6cm
\beqn
f_\alpha(x)= x f_\beta(x) = {\frac{p^2 x}{a^2 (1-p)+ \exp (- px /a )}}
\eeqn

\begin{figure}[h]
\centerline
{
\includegraphics[width=3.35cm,height=5.5cm,angle=270]{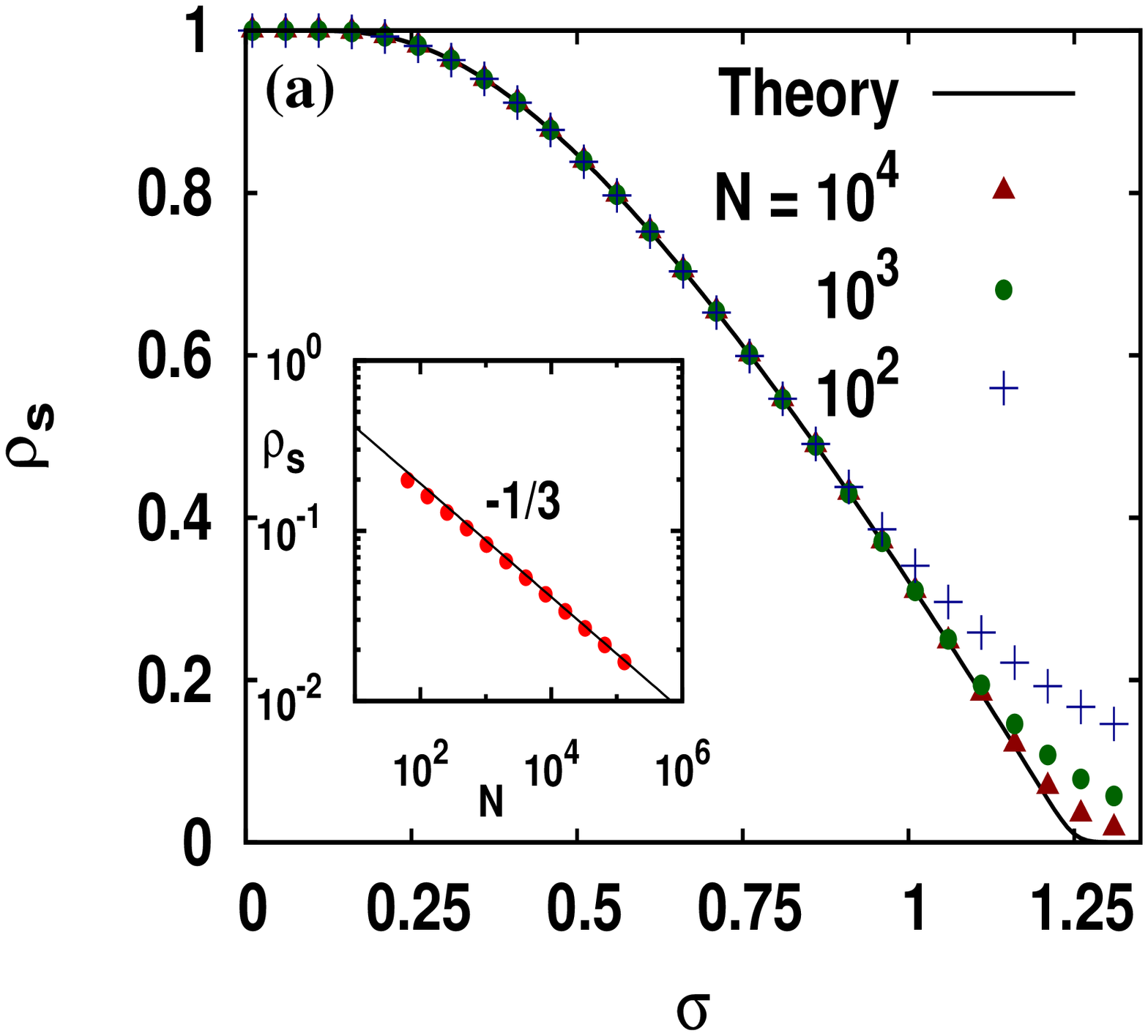}\hskip-0.25cm
\includegraphics[width=3.3cm,height= 5.0cm,angle=270]{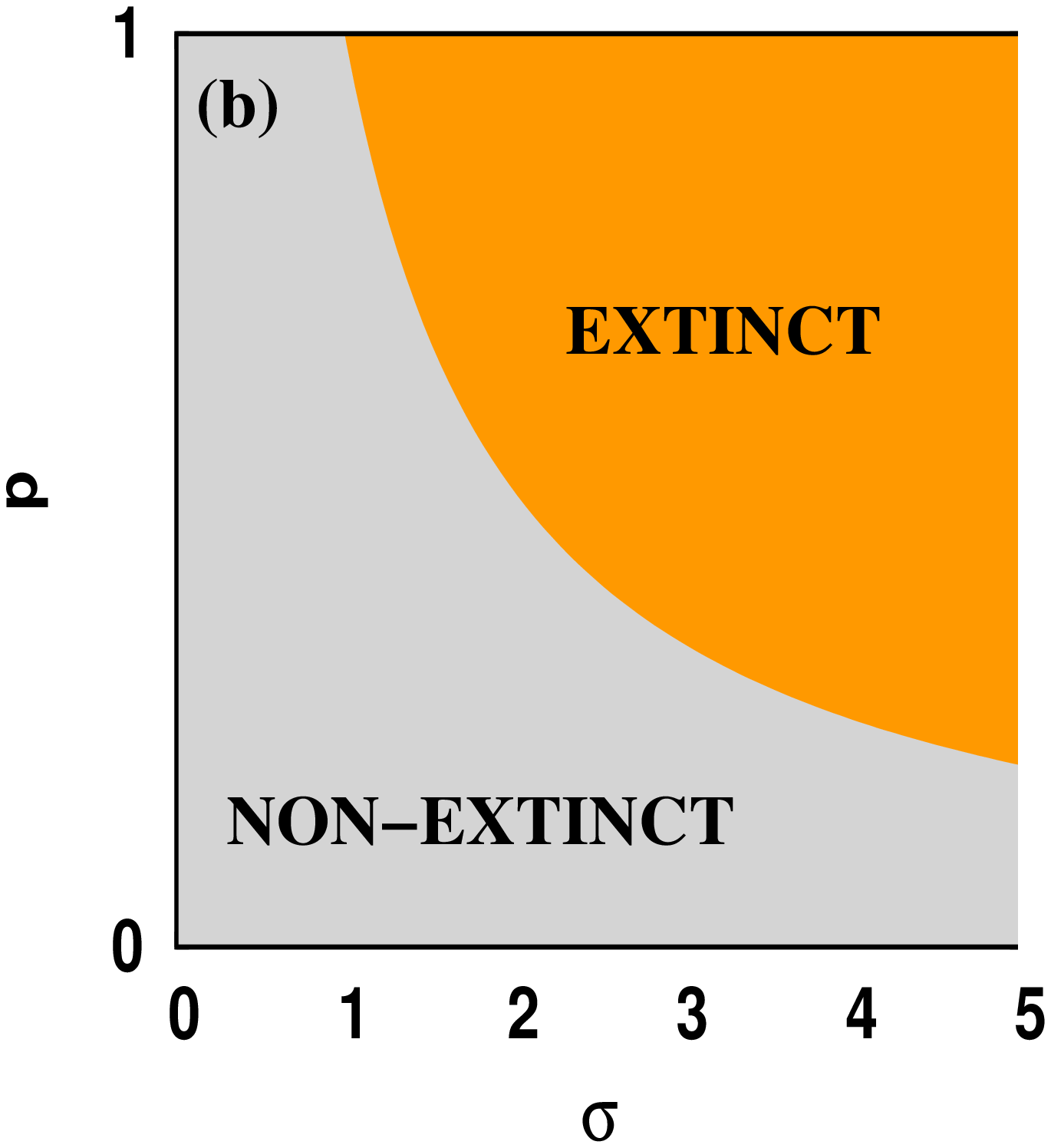}
}
\caption{(Color online) (a) $\rho_s$ as a function of   
$\sigma$ for  fitness  distribution ${\cal F} (x) = \exp (-1/x)/x^2$ and $p=0.8.$
Simulation results for different $N$(symbols) are compared 
with  solution of  Eq. (\ref{eq:fxd-pt}) (solid line).
(inset) log-log plot of $\rho_s$ {\it vs.} $N$  at $\sigma_c= 1/p$  is linear with
slope $-\beta/\nu_\perp=-1/3.$ 
(b) Phase diagram of the model; complete extinction occurs for $\sigma> \sigma_c=1/p.$ }
\label{fig:pt-beta-nu}
\end{figure}

A finite size scaling relation can also be derived for this model.
For a system of size $N$, the mean number of existing species 
$\langle N_t\rangle $ will have a finite nonzero value even
at the critical point $\sigma_c$ (see Fig.~\ref{fig:pt-beta-nu}), 
originating from  the  large critical fluctuations in $N_t.$ 
Fluctuations of $N_t$ result from the inherent stochasticity of 
the fitness values and  it is expected that its width will be 
proportional to $\sqrt{\langle N_t \rangle}.$ These fluctuations can 
be incorporated in the model  by  modifying the density $\rho_t$ as, 
\vskip-0.6cm
\beqn
\rho_t \rightarrow \rho_t + k \sqrt{\rho_t/N} 
\label{eq:fluc-N_t}
\eeqn
where, $\rho_t = \langle N_t \rangle /N $, is the mean density and $k$
is a constant of proportionality.  
Now using Eq.~(\ref{eq:fluc-N_t}) in the Eq.~(\ref{eq:fxd-pt-itr}) and  keeping 
the  leading terms in $N$ and $\rho$,  in the continuum limit we have,
$\dfrac{d\rho}{dt}  = k [ \sqrt{\rho/N}-K \rho^2],$ which on 
a rescaling of variables $\varrho  =  N^{1/3} \rho$  and ${\tt t} =  t N^{-1/3}$  yields, 
\vskip-0.6cm
\beqn
\dfrac{d\varrho}{d{\tt t}} = k  \left[  \sqrt{\varrho}-  K \varrho^2\right], 
\label{eq:fluc-N_t-2}
\eeqn
where $ K= {a(1-p)}/{pk}.$  The solution $\varrho({\tt t})$ can be rewritten\cite{note1}
in the scaling form
\beqn
\rho(N, t) &=& N^{-\beta/\nu_\perp} f_N(t N^{-z})
\eeqn
where  $z = \frac 13 = \beta/\nu_\perp$ (since $\beta =1$,  we have $\nu_\perp=3$). Note that the
dynamical exponent $z$ satisfy the  scaling relation $z = \nu_\shortparallel/ \nu_\perp.$ The 
unusually low  dynamical exponent  signifies  the {\it accelerated} extinction occurring in this model.

Until now, we have considered the system with infinite resources which 
however is unrealistic. 
For finite resources, the cooperativity function decreases with 
increase in the number of species $N_t,$ and a generic form can be chosen to be 
${\cal C } (N_t) = N_t^\theta \exp( {-N_t/N_m}),$
where  the scale $N_m$ is proportional to the total number of species $N.$ 
In this case, Eq.~(\ref{eq:evol-eqn-p}) with $p=1$ can be written as 
${\displaystyle \rho_{t+1} = 1- g\left(\frac{\sigma N^{1-\theta}}{ \rho_t^\theta \exp(-\rho_t/\rho_m)}\right)},$
where $\rho_m= N_m/N.$ Clearly, in the thermodynamic limit, this dynamics will
reach a stationary density $0<\rho_s<1$ only for $\theta =1.$ The
critical behavior in this case is identical to that of the system with infinite resources.

To demonstrate  the continuous  extinction transition we choose a specific 
fitness distribution ${\cal F} (x) = x^{-2} e^{-1/x}$ and a linear cooperativity function. 
The steady state density  for $p=1$ can be 
calculated exactly using Eq. (\ref{eq:fxd-pt}) as $\rho_s = 1 + \sigma W (-\sigma^{-1}\exp(-\sigma^{-1}))$                               
where $W(x)$ is the Lambert $W$ function. For $p\neq 1$, the solution of Eq. (\ref{eq:fxd-pt}) is found to be 
in excellent  agreement with $\rho_s$ obtained from numerical simulations  (Fig. \ref{fig:pt-beta-nu}(a)).
The inset shows that  $\rho_s\sim  N^{-\beta/\nu_\perp}$ with $\beta/\nu_\perp=1/3.$ 
The line of criticality  $\sigma= 1/p$ separates the  extinct phase from  non-extinct  one (Fig. \ref{fig:pt-beta-nu}(b)).  
This critical behavior  is robust against fluctuations  which has been checked  numerically  by   
adding gaussian noise to the environmental stress.

In conclusion, we have introduced an exactly solvable evolution model 
which incorporates some of the important features of biological evolution,
namely, extinction of less fit species under environmental stress,
cooperativity among the species and mutation. Under fairly general conditions,
the model undergoes a discontinuous phase transition into a fully extinct state, 
whereas for certain specific choice of fitness distribution and cooperativity
function the transition occurs continuously. The critical point, the critical exponents 
($\alpha, \beta, \nu_\shortparallel, \nu_\perp, z$)  and the scaling functions 
of the continuous transition are calculated analytically. In particular, the 
dynamical exponent $z = \frac 13$ is quite unexpected and implies an 
\textit{accelerated} species extinction. 
Apart from predicting large scale evolution $via$ a critical dynamics\cite{Sole} and 
existence of a completely extinct state, the model can also be applied to understand 
the trends in the available fossil data, as has been attempted by other models of
evolution\cite{review}. Close to criticality, the distribution of extinction events 
in this model follow a power-law with exponent $\tau= \frac32$ (obtained from  simulation of the model).
Since here the growth  exponent $\eta =0$ (no repopulation), one may argue\cite{munoz-tau} 
that $\tau = \frac{1+\eta+2\alpha}{1+\eta+\alpha} = \frac32$. 
In fact, this value is consistent with the distribution of extinct families of marine 
animal species, where $\tau$  varies in the range $1.35$ to $1.95$\cite{Sole}.
Our model can be extended further to include other complex features e.g., repopulation
of extinct species, local inter-species interactions in different dimensions,
species dependent external stress, speciation.

The authors thank M. Basu for providing help with the figures and
U. Basu for comments on the manuscript.


\begin{thebibliography}{99}
\bibitem{note1} Solution of Eq.~(\ref{eq:fluc-N_t-2}) is  ${\tt t} =  \frac{ 1}{3 k\sqrt[3]{K}} G(\sqrt[3]{K}\varrho),$ 
where $G(x) = 2 \sqrt{3} \tan ^{-1} (\frac{1+ 2 x}{\sqrt{3}}) + \log \frac{1+ x+ x^2}{(1-x)^2}$.
\end{thebibliography}
\end{document}